\documentstyle[epsfig,longtable]{aipproc}

\begin{document}
\title{Physics opportunities at RHIC and LHC}

\author{
S.~Scherer,
S.~A.~Bass$^*$, 
M.~Belkacem,
M.~Bleicher, 
J.~Brachmann, 
A.~Dumitru$^{\dagger}$, 
C.~Ernst, 
L.~Gerland,
N.~Hammon, 
M.~Hofmann,
J.~Konopka,
L.~Neise,
M.~Reiter,
S.~Schramm,
S.~Soff, 
C.~Spieles$^{\ddagger}$, 
H.~Weber, 
D.~Zschiesche, 
J.A.~Maruhn, 
H.~St\"ocker, 
W.~Greiner}
\address{
Institut f\"ur Theoretische Physik, Johann Wolfgang Goethe-Universit\"at\\ 
Robert Mayer-Str.\ 8--10\\ 
D-60054 Frankfurt am Main, Germany
\\
$^*$
present address: Department of Physics, Duke University, Durham, USA\\
$^{\dagger}$
present address: Physics Department, Yale University, New Haven, USA\\
$^{\ddagger}$
present address: Lawrence Berkeley Laboratory, Berkeley, USA\\
}

\maketitle

\begin{abstract}
Nonequilibrium models (three-fluid hydrodynamics, UrQMD, and quark
molecular dynamics) are used to discuss the uniqueness of often proposed 
experimental signatures for quark matter formation in relativistic heavy ion 
collisions from the SPS via RHIC to LHC. It is demonstrated that these models 
-- although they do treat the most interesting early phase of the collisions 
quite differently (thermalizing QGP vs.\ coherent color fields with virtual 
particles) -- 
all yield a reasonable agreement with a large variety  of the available 
heavy ion data.  Hadron/hyperon yields, including $J/\Psi$ meson  
production/suppression, strange matter formation, dileptons, 
and directed flow (bounce-off and squeeze-out) are investigated. 
Observations of interesting phenomena in dense matter are reported. 
However, we emphasize the need for systematic future measurements 
to search for simultaneous irregularities in the excitation functions 
of several observables in order to come close to pinning 
the properties of hot, dense QCD matter from data. The role of future 
experiments with the STAR and ALICE detectors is pointed out.
\end{abstract}

\section*{Introduction}

The study of relativistic  heavy ion 
collisions~\cite{gallmann2}
offers a unique chance to explore the properties of hot 
and dense elementary matter.
Throughout his scientific life, Klaus Kinder-Geiger has 
given new, creative and stimulating impulses to this exciting 
field of physics~\cite{geiger_qcd}. 
In the last years of his life, he had focused his attention on the 
very early phase of such collisions, when most of the energy is 
transfered to partonic degrees of freedom.

His parton cascade description \cite{geiger_vni} for the early 
stages of relativistic heavy ion collisions has become very important 
for our understanding of the experimental results gathered 
at the SPS at CERN \cite{geiger,geiger_sps}.
It will be crucial to the interpretation of data which will 
be collected at the STAR detector at RHIC 
\cite{geiger_longacre,geiger_star}.

RHIC will begin operation in 1999 with four detectors:
two medium scale ones, BRAHMS and PHOBOS, as well as two large scale 
detectors, PHENIX and STAR. The main emphasis of the STAR 
(Solenoidal Tracker At RHIC) detector will be the correlation 
of many (predominantly hadronic) observables 
on an event-by-event basis.

The great energy range and beam target range accessible
with RHIC will allow a dedicated systematic search
for the quark-gluon phase matter at energy densities
an order of magnitude above the transition domain.
This occurs not only because the rapidity density
of hadrons  is expected to be 2--4 times larger than in central SPS
collision, but also because   pQCD dominated mini-jet
initial conditions are finally reached in the collider 
($\sqrt{s}\sim 200$ AGeV) energy range. A whole class of 
new signatures involving hard pQCD probes (high $p_{\rm t}$ and jets) 
becomes available.

At yet higher energies at LHC, quark-gluon plasma research efforts and 
planning are centered around the ALICE detector. Its design is similar 
to that of STAR. ALICE will be the only large scale heavy ion detector 
setup at LHC.
At $\sqrt{s} \sim 5$ ATeV even bottom quarkonia are copiously produced
and transverse momenta twice as high ($p_{\rm t} \sim 60$ GeV/c)
will be readily measurable to probe even deeper into the multiparticle
dynamics of a QGP.

\section*{Critical review of QGP signatures}

In the last few years researchers at Brookhaven and CERN have 
succeeded to measure a wide spectrum of observables with 
heavy ion beams, $Au+Au$ and $Pb+Pb$. While these programs 
continue to measure with greater precision the beam energy--, 
nuclear size--, and centrality dependence of those observables, 
it is important to recognize the major milestones for 
relativistic heavy ion physics passed thus far in that work. 

The experiments have conclusively demonstrated the existence of strong nuclear 
$A$ dependence of, among others, $J/\psi$ and $\psi'$ meson 
production and suppression, strangeness enhancement, hadronic resonance 
production, stopping and directed collective transverse and longitudinal 
flow of baryons and mesons -- in and out of the impact plane, 
both at AGS and SPS energies --, and 
dilepton-enhancement below and above the $\rho$ meson mass. 
These observations support that a novel form of ``resonance matter'' 
at high energy- and baryon density has been created in nuclear collisions.
The global multiplicity and transverse energy measurements prove 
that substantially more entropy is produced in $A+A$ collisions at the SPS
than simple superposition of $A\times pp$ would imply. 
Multiple initial and final state interactions play a critical role 
in all observables. 
The high midrapidity baryon density (stopping) and the observed 
collective transverse and directed flow patterns constitute one
of the strongest evidence for the existence of an extended period 
($\Delta \tau\approx 10$~fm/c) of high pressure and 
strong final state interactions. The enhanced $\psi'$ 
suppression in $S+U$ relative to $p+A$ also attests to this fact. 
The anomalous low mass dilepton enhancement shows that substantial 
in-medium modifications of multiple collision dynamics exists, 
probably related to in-medium collisional broadening of vector mesons. 
The non-saturation of the strangeness (and anti-strangeness) 
production shows that novel non-equilibrium production 
processes arise in these reactions. 
Finally, the centrality dependence of $J/\psi$ absorption 
in $Pb+Pb$ collisions presents further hints towards the nonequilibrium 
nature of such reactions. 
Is there evidence for the long sought-after quark-gluon plasma 
that thus far has only existed as a binary array of predictions 
inside teraflop computers?

As we will discuss, it is too early to tell. 
Theoretically there are still too many ``scenarios'' and 
idealizations to provide a satisfactory answer. 
Recent results from microscopic transport models as well as 
macroscopic hydrodynamical calculations differ significantly 
from predictions of simple thermal models, e.~g.\ in the flow pattern. 
Still, these nonequilibrium models provide reasonable predictions 
for the experimental data. We may therefore be forced to rethink our
concept of what constitutes the deconfined phase in ultrarelativistic 
heavy-ion collisions. Most probably it is not a blob of thermalized quarks and
gluons. Hence, a quark-gluon plasma can only be the 
source of {\em differences} to the predictions of these models for 
hadron ratios, the $J/\Psi$ meson production, dilepton yields, 
or the excitation function of transverse flow.
And there are experimental gaps such as the lack of 
intermediate mass $A\approx 100$ data and the limited number of beam energies 
studied thus far, in particular between the AGS and SPS. 

In the future, the field is at the doorstep of the next milestone: 
$A+A$ at $\sqrt{s}=30-200$ AGeV are due to begin at RHIC/BNL 
in the summer of 1999,
and at even higher energies ($\sqrt{s}<5$ ATeV) at LHC/CERN in the
next millennium. 

\begin{figure}[tb] 
\centerline{\epsfig{file=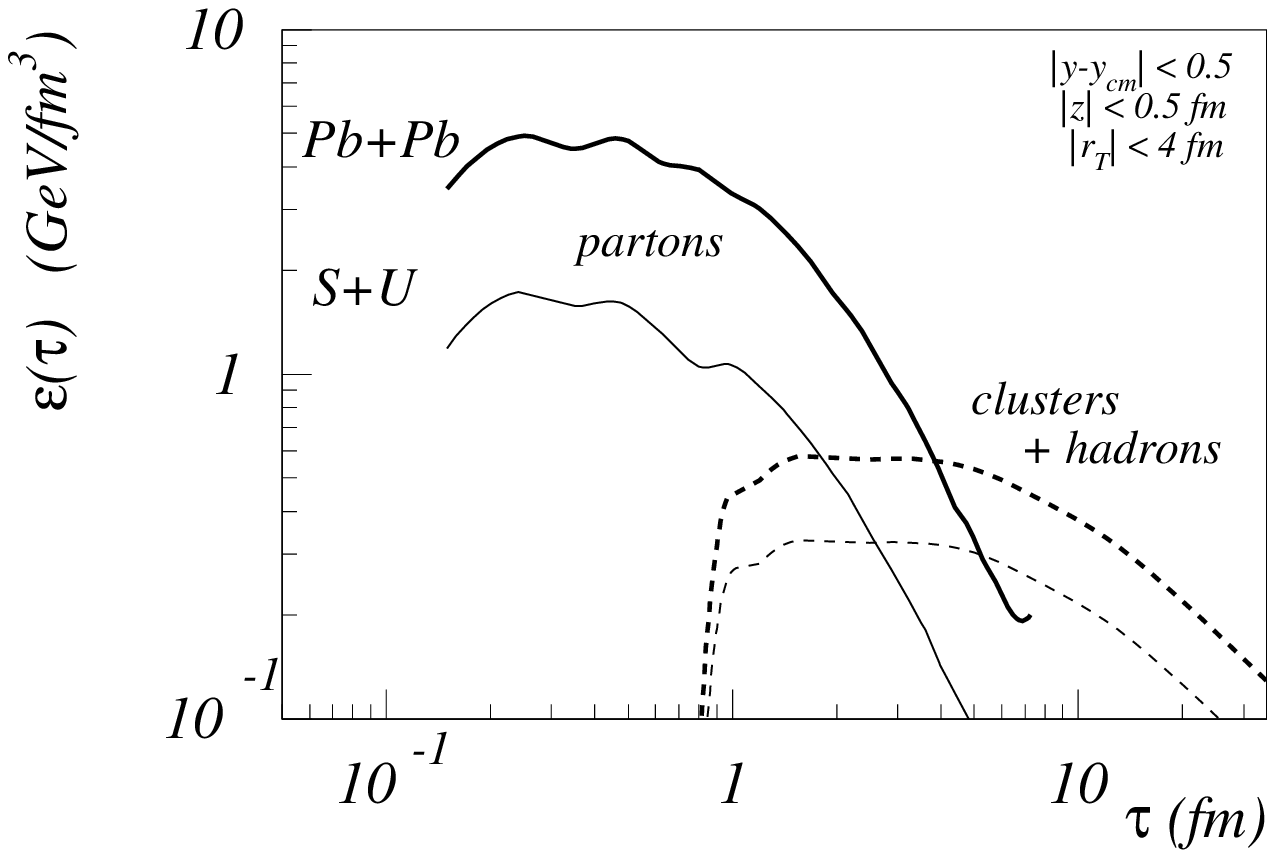,width=8cm}
\hfill 
\epsfig{file=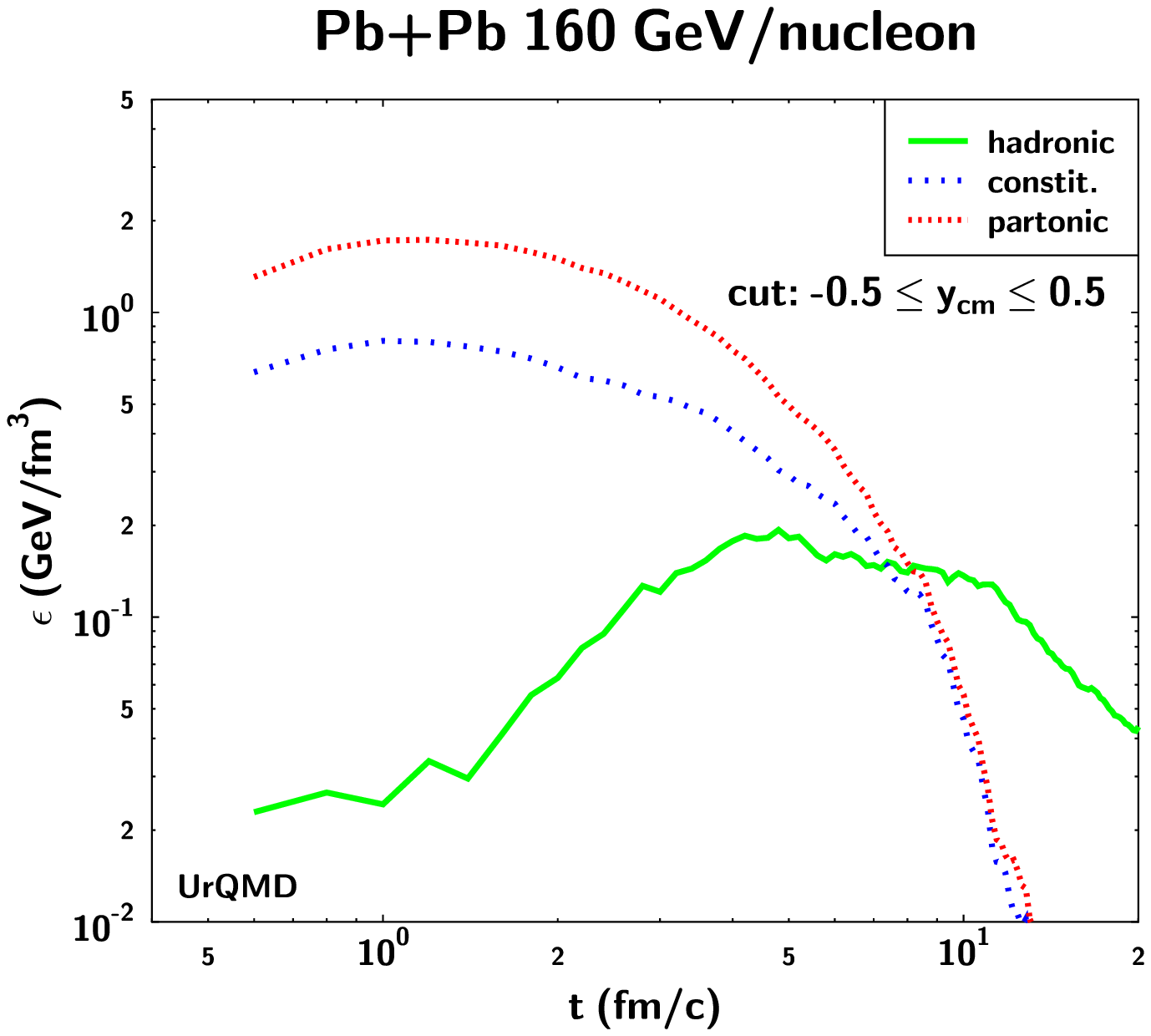,width=8cm}}
\vspace{10pt}
\caption{
(Left) Klaus Kinder-Geigers parton cascade (VNI, from \protect\cite{geiger}) 
and (Right) UrQMD  results for the time 
evolution of energy density in central $Pb+Pb$ reactions at 160 AGeV. 
At an early stage, most of the energy is contained 
in the partonic degrees of freedom (VNI) or in constituents (UrQMD).}
\label{pcm}
\end{figure}

Here, the results of Klaus Kinder-Geiger are of great importance for
an understanding of the data to come: Any theoretical description of the 
early stages of events observed with these machines, reaching extremely 
high energy and matter densities, must take care of the partonic degrees 
of freedom (see \cite{geiger}, and Fig.~\ref{pcm}). One way of doing this is 
the use of parton cascade models, as initiated and furthered by Klaus.

\section*{Nonequilibrium models}

In the present survey of relativistic heavy ion collisions,  
we employ two sharply distinct nonequilibrium models,
namely the macroscopic 3-fluid hydrodynamical model \cite{hydro_review}
and the Ultra-relativistic Quantum Molecular Dynamical model, 
UrQMD \cite{urqmd_review}. 
The first model assumes that a projectile- and a target fluid 
interpenetrate upon impact of the two nuclei, 
creating a third fluid (in the present version baryon free, 
see, however, \cite{rosenhauer}) 
via new source terms in the continuity equations 
for energy- and momentum flux. 
Those source terms are taken from energy- and rapidity 
loss measurements in high energy $pp$-collisions. 
The equation of state (EoS) of this model assumes equilibrium only 
in each fluid separately and allows for a first order phase transition 
to a quark gluon plasma in fluid 1, 2 or 3, 
if the energy density in the fluid under consideration 
exceeds the critical value for two phase coexistence. 
Pure QGP can also be formed in every fluid separately, 
if the energy density in that fluid exceeds the maximum energy 
density for the mixed phase.
The UrQMD model, on the other hand, assumes an independent 
evolution of hadrons, strings, and constituent quarks and diquarks 
in a nonequilibrium multiparticle system. 
The collision terms in this system of coupled Boltzmann 
(partial differential-/integral-) equations are taken from 
experimental data, where available, and otherwise from additive quark model 
and string phenomenology.

What is the role of partonic degrees of freedom in relativistic 
heavy ion reactions at the SPS?

\noindent
Fig.~\ref{pcm} shows the time evolution of the energy density $\epsilon$ 
in central $Pb+Pb$ reactions at 160 AGeV as obtained within
{\em a)} the parton cascade approach VNI \cite{geiger} of Klaus 
Kinder-Geiger, {\em b)} the UrQMD model \cite{weber98}. 
It can be seen that in both models and at early times of the collision, 
a large fraction of the energy density is contained in partonic
degrees of freedom (VNI) or  to nearly equal parts in constituent diquarks
and quarks from the strings and in virtual hadrons.
This (virtual) ``partonic'' phase in $Pb+Pb$ reactions at 160 AGeV is,
however, not to be identified with an equilibrated QGP.
Note that the absolute values differ by a factor 2 in the two models and 
depend heavily on the rapidity cuts imposed to discriminate between
virtual free streaming and interacting matter.

\begin{figure}[tb] 
\centerline{\epsfig{file=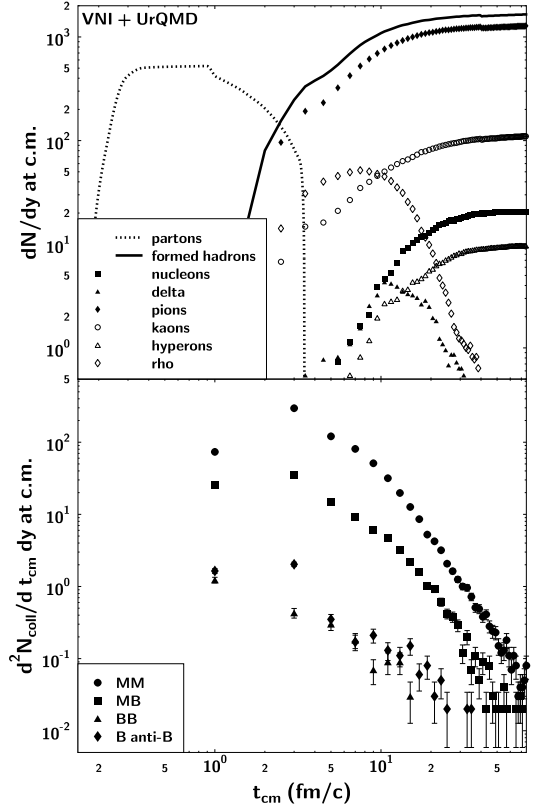,width=7.5cm}
} 
\vspace{10pt}
\caption{ 
(Top) Time evolution of the parton and on-shell hadron
rapidity densities at c.m. for central ($b\le 1$~fm) $Au+Au$ collisions at
RHIC. There exists a considerable overlap
between the partonic and hadronic phases of the reaction. Hadronic resonances
are formed and remain populated up to $\approx$~20~fm/c indicating a large
amount of hadronic interaction. (Bottom) Rates for hadron-hadron collisions per
rapidity at c.m.. Meson-meson and to a lesser extent meson-baryon
interactions dominate the dynamics of the hadronic phase.
} 
\label{tevol}
\end{figure}

While there is currently a strong debate whether
(equilibrated) deconfined matter may have been produced at the CERN/SPS, 
it is widely expected that in collisions of heavy nuclei at 
RHIC a QGP will be formed.
The consequence is that at RHIC, both, partonic and hadronic, degrees
of freedom have to be treated explicitly. A large step in that direction
has been recently undertaken by modeling the initial parton dynamics 
in the framework of the parton cascade model, performing hadronization
via a cluster hadronization model and configuration space
coalescence, and finally describing the hadronic phase either by a hadronic
after-burner \cite{geiger_longacre} or by a full microscopic
hadronic transport approach \cite{bass_vni1}. 
Technically the latter approach
is realized by combining VNI \cite{geiger} for the initial phase and 
hadronization with UrQMD \cite{urqmd_review}
for the later, hadronic, reaction stages. 
The resulting reaction dynamics indicates
a strong influence of hadronic rescattering on the space-time pattern
of hadronic freeze-out and on the shape of transverse mass spectra.
The upper frame of figure~\ref{tevol} shows the time evolution (in c.m. time, 
$t_{c.m.}$)
of the rapidity density dN/dy of partons (i.e. quarks and gluons) 
and on-shell hadron
multiplicities at $|y_{c.m.}|\le 0.5$.
Note that there are no distinctly separate time scales for the three
reactions stages discussed earlier in this article: 
hadronic and partonic phases may evolve in
parallel and both, parton-parton as well as hadron-hadron interactions 
occur in the same space-time volume.
The overlap between the partonic and hadronic stages of the reaction 
stretches from $t_{c.m.} \approx 1$~fm/c up to $t_{c.m.} \approx 4$~fm/c
for the midrapidity region. The calculation indicates 
that this overlap occurs 
not only in time but also in coordinate space -- partonic and hadronic degrees
of freedom occupy the same space-time volume during this reaction 
phase \cite{bass_vni1}.
Hadronic resonances like the $\Delta(1232)$ and the $\rho(770)$ (which
are the most abundantly produced baryonic and mesonic resonance states) 
are formed and remain populated up to 
$t_{c.m.} \approx 15 - 20$~fm/c, indicating a considerable
amount of hadronic rescattering. Hadron yields saturate
at time-scales $t_{c.m.} \approx 25$~fm/c. 
Since resonance decays have not been factored into this estimate
of the saturation time, this number should be viewed as an upper estimate for 
the time of chemical freeze-out.

Rates for hadron-hadron collisions
per unit rapidity at $y_{c.m.}$ 
are shown in the lower frame of figure~\ref{tevol},
i.e. all hadron-hadron collisions for hadrons with $|y_{c.m.}| \le 0.5$
were taken into account. 
Meson-meson and  meson-baryon
interactions dominate the dynamics of the hadronic phase. Due to their
larger cross sections baryon-antibaryon collisions occur more
frequently than baryon-baryon interactions. However, both are suppressed
as compared to meson-meson and meson-baryon interactions.
This is due to the large meson
multiplicity, which creates a  ``mesonic medium'' in which the
baryons propagate.

A comparison of calculations with and without hadronic rescattering shows that
e.g. the proton and antiproton multiplicities change by a factor of two due to 
hadronic rescattering, whereas the ratio of their yields remains roughly 
constant. Evidently chemical freeze-out of the system occurs well into the
hadronic phase and not at the ``phase-boundary''. The collision rates indicate
that interactions cease at $t_{c.m.} \approx 30-40$~fm/c 
at which point the system can be regarded as kinetically frozen out. Since
the saturation of the hadron yields occurs earlier, there is a clear separation
between chemical and kinetic freeze-out.

\section*{Yields of Hadronic Probes}

Let us now discuss the results obtained from hadronic probes, 
such as observed production of $J/\Psi$ mesons, enhancement of 
strange baryons, light mesons, and particle ratios. 
Observed hadrons include feeding by the decay of resonances.

%
%
\subsection*{$J/\psi$ suppression}

\begin{table}[t!bh] 
\begin{tabular}{|c|c|c|c|c|c|c|c|c|}
\hline
$c\overline{c}$/$b\overline{b}$-state & J/$\Psi$ & $\Psi'$ & $\chi_{c10}$ & $\chi_{c11}$\\
\hline
$<b^2>$ (fm${}^2$) & 0.094 & 0.385 & 0.147 & 0.293 \\
\hline
$\sigma_{\rm nonperturbative}$ (mb) & 3.62 & 20.0 & 6.82 & 15.9 \\
\hline
$\sigma_{\rm hard}$ (mb) (SPS)& 0.024 & 0.012 & 0.021 & 0.006 \\
\hline
$\sigma_{\rm hard}$ (mb) (RHIC)& 1.73 & 0.68 & 1.23 & 0.30 \\
\hline$
\sigma_{\rm hard}$ (mb) (LHC)& 20.8 & 8.2 & 14.7 & 3.5 \\
\hline
\end{tabular}  
\vspace{10pt}
\caption{\label{meanb}
The average square of the transverse distances of the charmonium
states and the total quarkonium-nucleon cross sections $\sigma$.
For the $\chi$ two values arise, due to the spin dependent wave functions
($lm=10,11$). $\sigma_{\rm hard}$ are the perturbative QCD contribution
at different energies.}
\end{table}

Debye screening of heavy charmonium mesons in an equilibrated  
quark-gluon plasma may reduce the range of the attractive force 
between heavy quarks and antiquarks \cite{matsui_satz}.
Mott transitions then dissolve particular bound states, one by one.  
NA38 found evidence of charmonium suppression in light ion reactions. 
Then also in $p+A$ such suppression was observed. 
New preliminary $Pb+Pb$ data of NA50 show ``anomalous'' suppression.

One of the main problems in the interpretation of the observed suppression 
as a signal for deconfinement is 
that non-equilibrium dynamical sources of charmonium suppression 
have also been clearly discovered in $p+A$ reactions, 
where the formation of an equilibrated quark-gluon plasma 
is not expected.
A recent development is the calculation of the hard contributions 
to the charmonium- and bottonium-nucleon cross sections based 
on the QCD factorization theorem and  the non-relativistic 
quarkonium model \cite{gerland98a}. 
Including non-perturbative contributions,
the calculated $p+A$ cross section agrees well with the data.

\begin{figure}[tbh]
\centerline{\epsfig{file=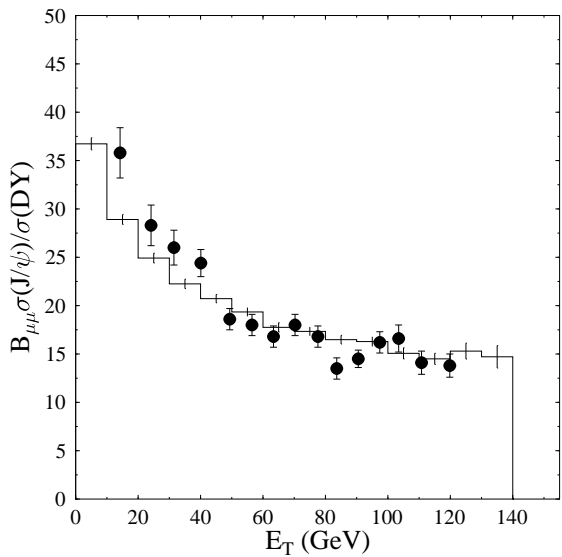,width=8cm}} 
\vspace{10pt}
\caption{
The ratio of $J/\psi$ to Drell-Yan production as a function 
of $E_{\rm t}$ for $Pb+Pb$ at 160~GeV. The experimental data are from 
Ref.~\protect\cite{romana}, the histogram is a UrQMD 
calculation~\protect\cite{spieles98b}. No scaling factor has been 
applied to the $x$-axis for either the calculations or the data.} 
\label{jpsi_dyet}
\end{figure}

The numerical calculation shown in Tab.~\ref{meanb} was done for 
charmonium states produced in midrapidity at SPS energy and 
in the target fragmentation region at RHIC and LHC. 
One can see that the hard contribution to the cross section is just a
correction at SPS energies, but at RHIC energies both contributions
become comparable and at LHC it dominates (one neglects here that the DGLAP
equation (Dokshitser-Gribov-Lipatov-Altarelli-Parisi) should be probably
violated~\cite{felix}).

Whereas these descriptions of nuclear absorption can account for the 
$p+A$ observation, the corrections needed for an extrapolation to $A+A$ 
reactions are, however, not yet under theoretical control.

Purely hadronic dissociation scenarios have been 
suggested \cite{neubauer88a,gavin90,gavin97} 
which could account for $J/\psi$ and $\psi'$ 
suppression without invoking the concept of deconfinement
(``comover models'').  Suppression in excess to that 
due to preformation and nuclear absorption is ascribed in such models 
to interactions of the charmonium mesons with ``comoving'', but 
probably off-equilibrium, mesons and baryons, 
which are produced copiously in nuclear collisions. 
Fig.~\ref{jpsi_dyet} shows an UrQMD calculation 
which employs a microscopic free streaming 
simulation for $J/\psi$ production and a microscopic transport 
calculation for nuclear and comover dynamics as well as for 
rescattering \cite{spieles98b}. The dissociation cross sections 
are calculated using the QCD factorization theorem \cite{gerland98a}, 
feeding from $\psi'$ and $\chi$ states is taken into account, 
and the $c \bar c$ dissociation cross sections increase linearly 
with time during the formation of the charmonium state. 
Taking into account the non-equilibrium ``comovers''   
($\sigma_{\rm meson} \approx 2/3 \sigma_{\rm nucleon}$), 
the agreement between theory and data is reasonable (Fig.~\ref{jpsi_dyet}). 
New, unpublished data agree better with the model predictions, 
but the high and low $E_{\rm t}$ regions remain to be studied carefully 
in the experiment. At present, no ab initio calculation does predict 
sudden changes in the suppression. In fact, from three-fluid calculations, 
even with QGP phase included, only a moderate change 
of the average and local energy density with bombarding energy is predicted.  
This seems to strongly speak against drastic threshold 
effects in the charmonium production.

The strong dependence of these results on details, such as the treatment 
of the formation time or the time dependent dissociation cross section, 
remain to be studied further. 
Furthermore, quantum effects such as energy dependent formation and 
coherence lengths must be taken into account \cite{huefner96a} 
before definite statements can be made with regard to the nature of 
the $J/\psi$ suppression. 
Interpretations of the data based on plasma scenarios are also 
increasingly evolving away from the original Mott transition analog
\cite{kharzeev96,satz98}.

Hence, the theoretical debate on the interpretation of the pattern 
of charmonium suppression discovered by NA38/NA50 at the SPS is far 
from settled. It is not clear whether the suppression is 
the smoking gun of nonequilibrium dynamics or deconfinement. 
It is not likely to be due to simple Debye screening.

The major goal of further theoretical work is not to continue 
to try to rule out more ``conventional'' explanations, 
but to give positive proof of additional suppression by QCD-calculations
which actually {\em predict} the $E_{\rm t}$-dependence of the
conjectured signature. Consistency tests and a detailed simultaneous 
analysis of all other measured observables are needed, if at least 
the same standards as for the present calculations are to be hold up.

%
%
\subsection*{Particle ratios}

The study of particle ratios has recently attended great interest 
at the AGS \cite{ogloblin,satz,heinz} and at the SPS 
\cite{PBMAGS,braunmunz96}.

One assumes a thermalized system with a constant density 
$\rho(r)$ (box profile), a constant temperature $T(r)$ and a linear radial 
and longitudinal flow velocity profile $\beta_{\perp}(r)$, $\beta_{||}(r)$. 
These parameters are assumed to be the same for all hadrons/fragments.
At some time $t^{\rm break-up}$ and density $\rho^{\rm break-up}$, 
the system decouples as a whole (a horizontal freeze-out in the
$T(z)$-plane) and the particles are emitted instantaneously 
from the whole volume  of the thermal source. 
A complete loss of memory results, due to thermalization --
the emitted particles carry no information about the evolution of the source.
If one wants to use the inverse slope parameter~$T$ as 
thermometer \cite{stoecker86a}, the feeding from $\Delta$'s etc., as well 
as the radial flow need to be incorporated into the analysis. The same
holds for the use of d/p, $\pi$/p etc. as an entropymeter\cite{siemens}.
In addition, the proper Hagedorn volume correction can be applied
\cite{sigom}.
A two parameter fit ($\mu_{\rm q}$, T, $\mu_{\rm s}$ is 
fixed by strangeness conservation) to the hadronic freeze-out data 
describes the experimental results well, if feeding is included 
 \cite{braunmunz96}. 
Does this compatibility with a thermal source proof volume emission from a 
globally equilibrated source?

\begin{figure}[tb]
\centerline{\epsfig{file=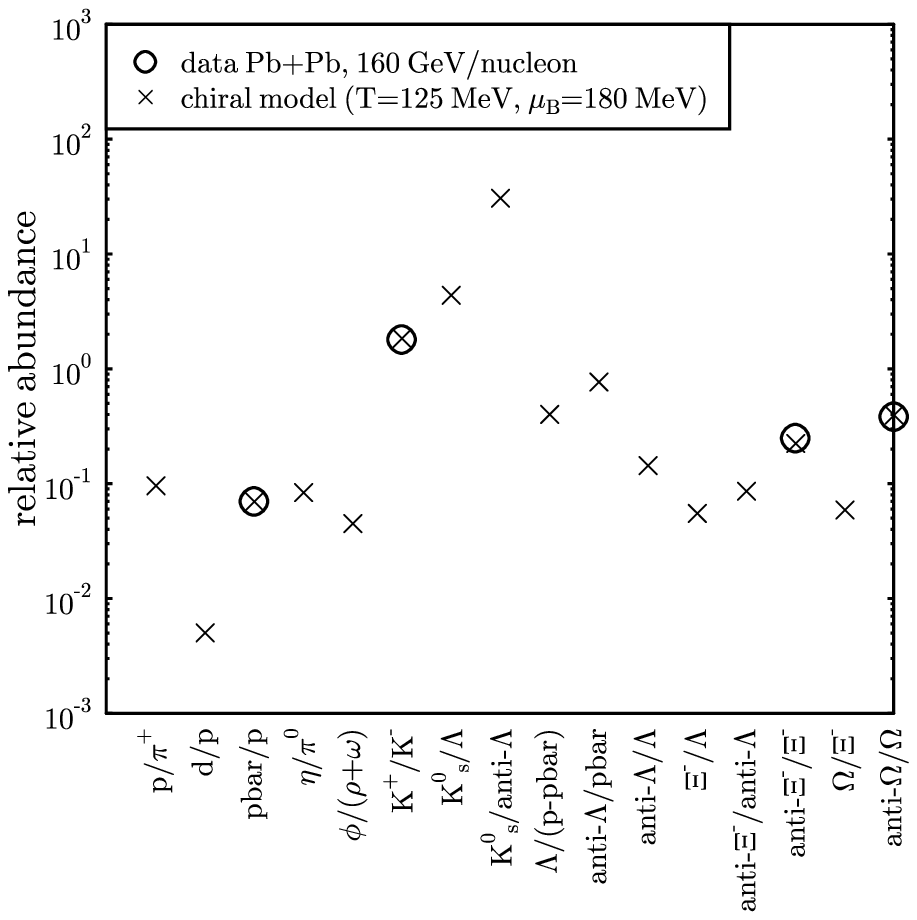,width=8cm}
\hfill 
\epsfig{file=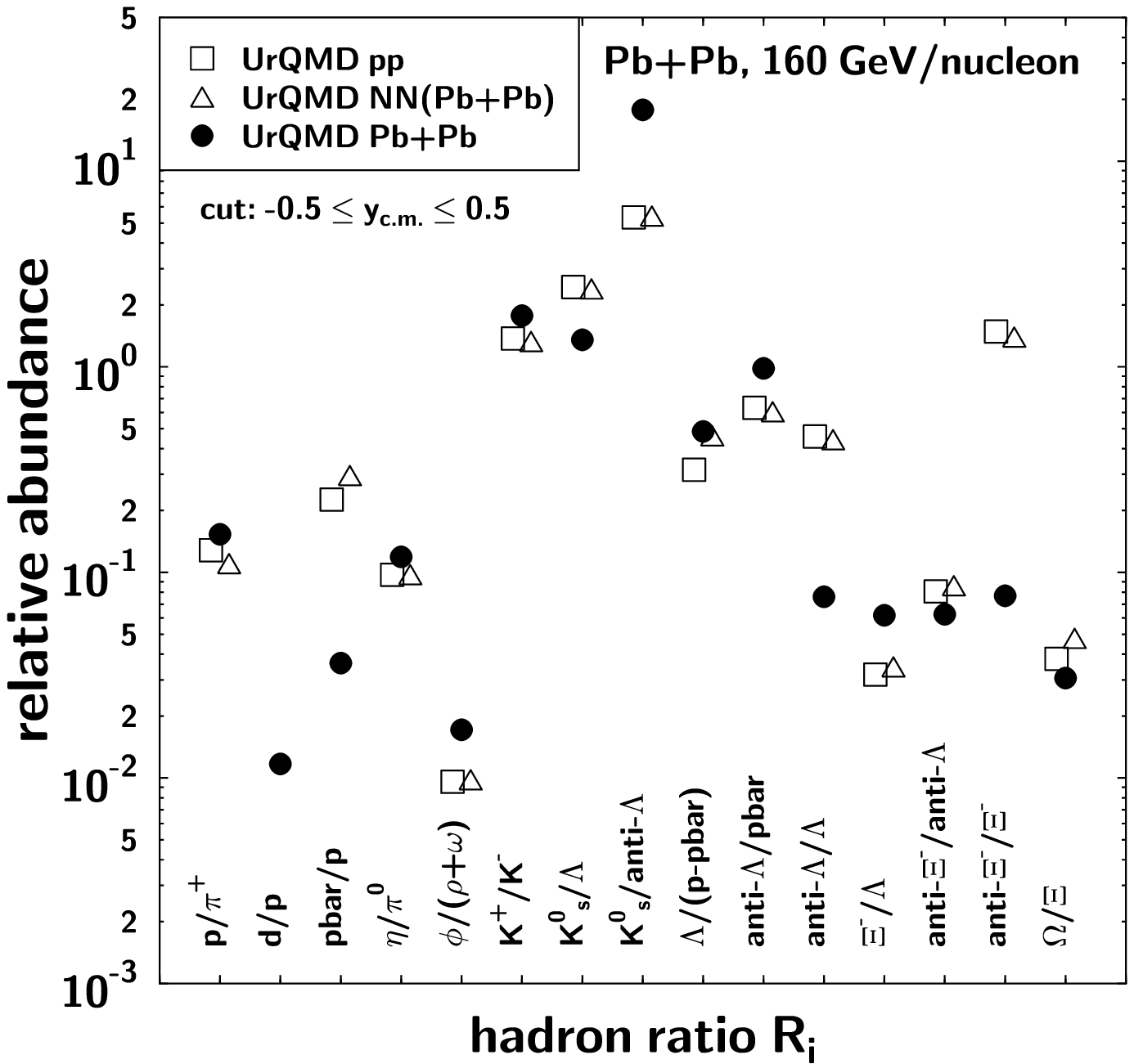,width=8cm}}
\vspace{10pt}
\caption{
(Left) Fit of hadron ratios 
from the chiral model to preliminary data from $Pb+Pb$ collisions at SPS. 
The obtained values of $T$ and $\mu$ allow
the prediction of further ratios. $T$ and $\mu$ are much lower than thermal model
results from the ideal hadron gas . 
(Right) UrQMD prediction for hadron ratios in $Pb+Pb$ collisions 
at midrapidity (full circles), compared to a superposition 
of pp, pn and nn reactions with the isospin weight of the $Pb+Pb$ system 
(open triangles), i.e. a first collision approach.} 
\label{ratios_urqmd_chiral} 
\end{figure}

The ideal gas thermal fit to experimental data for 
hadron ratios in $S+Au$ collisions at 200 AGeV  
gives values for the parameters $T$ and $\mu_{\rm B}$ which can be used as 
input for a $SU(3)$ chiral mean-field model \cite{papa98} extended to 
finite temperatures \cite{zsch98}. Feeding from the decay of higher 
resonances is included. 
One finds that in such a model (which selfconsistently contains 
a chiral phase transition at $T \approx 150$ MeV) the ideal gas model 
values $T=160$ MeV and $\mu_{\rm B}=170$~MeV lead to strong deviations 
from the experimental data.  Only the $\Omega/\Xi^-$-ratio is in a 
good agreement, in contrast to the ideal gas model. 
Hence, the system can not be close to the chiral phase transition -- 
the $T$ and $\mu$ values extracted from the free thermal model cannot 
be identified with the real temperature and chemical potential of the system!

The chiral mean-field model does reproduce the
data compiled in \cite{gor98} for relative abundances in 
$Pb+Pb$ collisions at 160 AGeV (Fig.~\ref{ratios_urqmd_chiral}, left) 
for $T=125$ MeV and $\mu_B= 180$, much lower than the thermal model 
results \cite{braunmunz98,gor98} ($T=160-175$~MeV, $\mu_{\rm B}=200-270$ MeV).
The microscopic UrQMD transport model is in good agreement with the measured hadron ratios 
of the system $S+Au$ at CERN/SPS \cite{bass98a}. A thermal model fit to 
the calculated ratios  yields a temperature of $T=145$ MeV and a chemical 
potential of $\mu_{\rm B}=165$~MeV. However, these ratios exhibit a strong 
rapidity dependence. Thus, thermal model fits to data may be distorted 
due to different acceptances for the individual ratios. 

Hadron ratios for the system $Pb+Pb$ are predicted by UrQMD and can be 
fitted by a thermal model with $T=140$~MeV and $\mu_{\rm B}=210$~MeV 
(Fig.~\ref{ratios_urqmd_chiral}, right). Analyzing the results of 
non-equilibrium transport model calculations by an equilibrium model may,
however, be not meaningful.

There is a problem in the definition of equilibrium in itself: 
Do heavy ion collisions ever reach a thermalized system? 
Or are there transient steady states off equilibrium \cite{zabrodin}? 
Due to the rapid dynamics of the system, the assumption of 
detailed balance is not fulfilled in the initial stage. 
This drives the system into a steady state far from equilibrium, 
but stationary in time. This steady state is easily visible 
in an enhanced production of light mesons, 
as compared to thermal models.

During the initial off-equilibrium stage of energetic nuclear collisions, 
a large amount of entropy can be produced \cite{entropy}. The subsequent 
expansion is, on the other hand, often assumed to be nearly isentropic. 
The entropy produced during the compression stage is closely linked to 
the finally observable relative particle yields.

This entropy production can be calculated \cite{reiter98a} within 
three-fluid hydrodynamics. The entropy per net participating baryon, $S/A$, 
saturates rapidly as a function of CM-time and is essentially time 
independent for later times when the freeze-out is reached. 
The chemical composition of the fireball is given by the net baryon 
density, the net (zero) strangeness of the system, 
and the specific entropy $S/A$, as described for the thermal model above.

\begin{figure}[tb]
\centerline{\hbox{\epsfig{figure=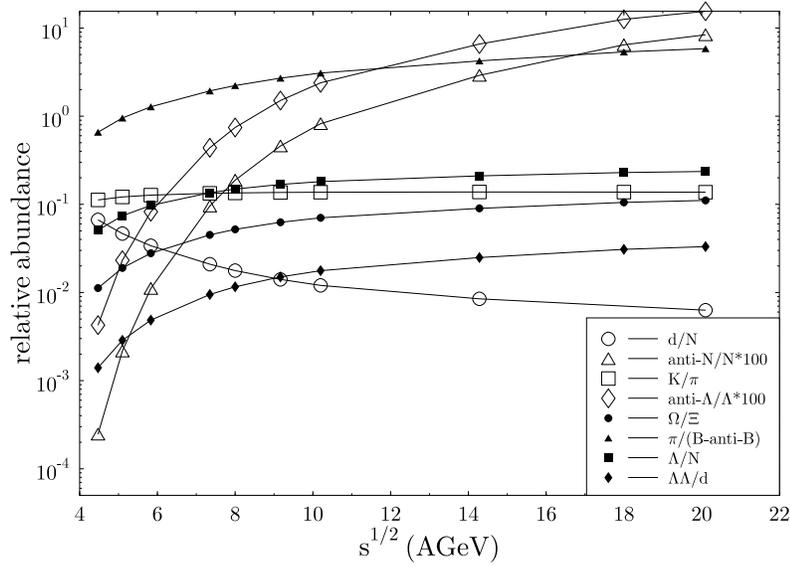,width=11cm}}} 
\vspace{10pt}
\caption{
The excitation function of various particle ratios as calculated 
from the S/A values obtained from the three-fluid model. 
Feeding due to decays of resonances is taken into account.} 
\label{fig_ratiosws}
\end{figure}

The hadron ratios thus obtained are shown in Fig.~\ref{fig_ratiosws}. 
At AGS and SPS energies, they are quite close to the data 
\cite{PBMAGS,bravina}. 
For such a simple estimate 
of hadron production in nuclear collisions, deviations from the 
experimental ratios by up to factors of two have to be expected. 
Nevertheless, it is clear from Fig.~\ref{fig_ratiosws} that 
the simultaneous measurement of various hadron ratios, 
like $\pi/\left(B-\overline{B}\right)$, d$/N$ and, in particular, 
$\overline{B}/B$ (provided antibaryons also reach chemical equilibrium) 
allows to determine the produced entropy in the energy range 
between the AGS and the SPS. In contrast, 
the $K/\pi$-ratio is practically constant.
The total specific entropy $S/A$ produced within the three-fluid model 
is consistent with the $S/A$ values extracted from data 
using relative particle yields from the thermal model. 
One finds $S/A=11$ for AGS and $S/A=38$ for SPS energies. 

The excitation function of the specific entropy $S/A(\sqrt{s})$ 
does not exhibit any threshold signatures of the phase transition to the QGP
incorporated in the EoS. This is due to the gradual transition through
the wide coexistence region in the energy density between 
$E_{\rm lab}\approx$~10--100~AGeV.

%
%

\subsection*{Strange baryons and mesons, Hypermatter and Strangelets}

\begin{figure}[tbh] 
\centerline{\epsfig{file=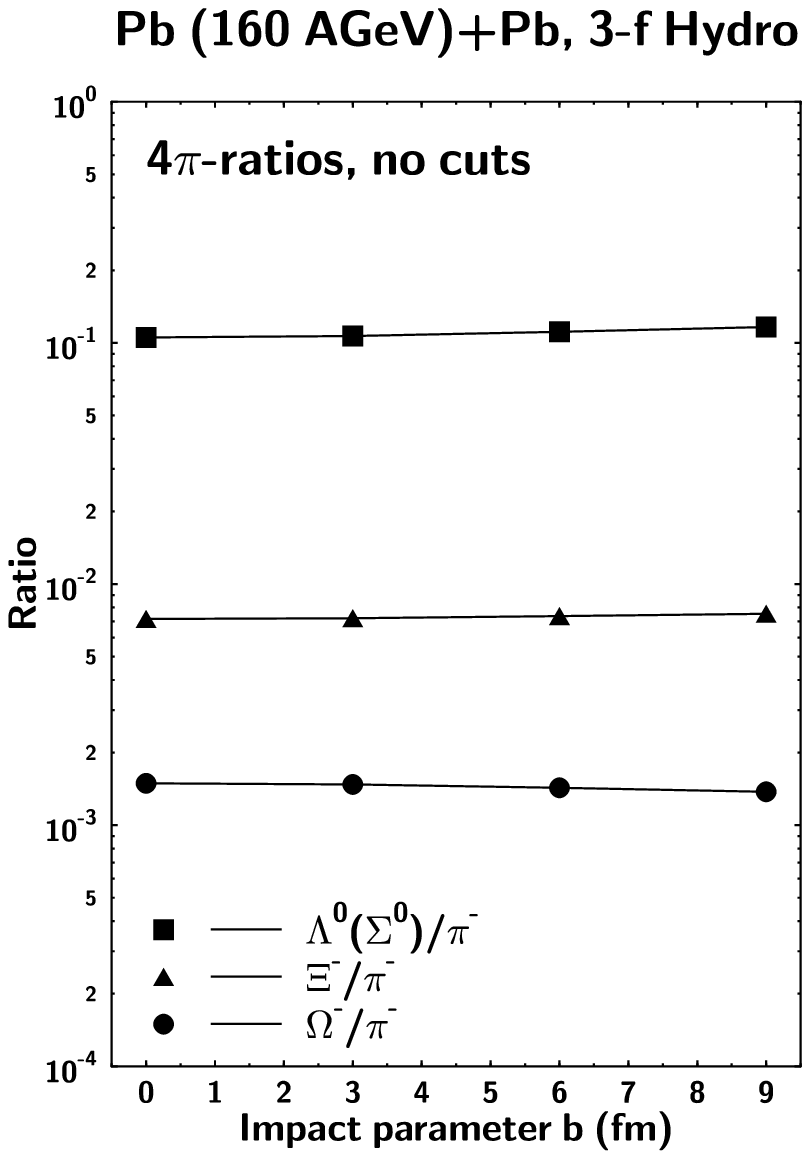,width=8cm}\hfill
\epsfig{file=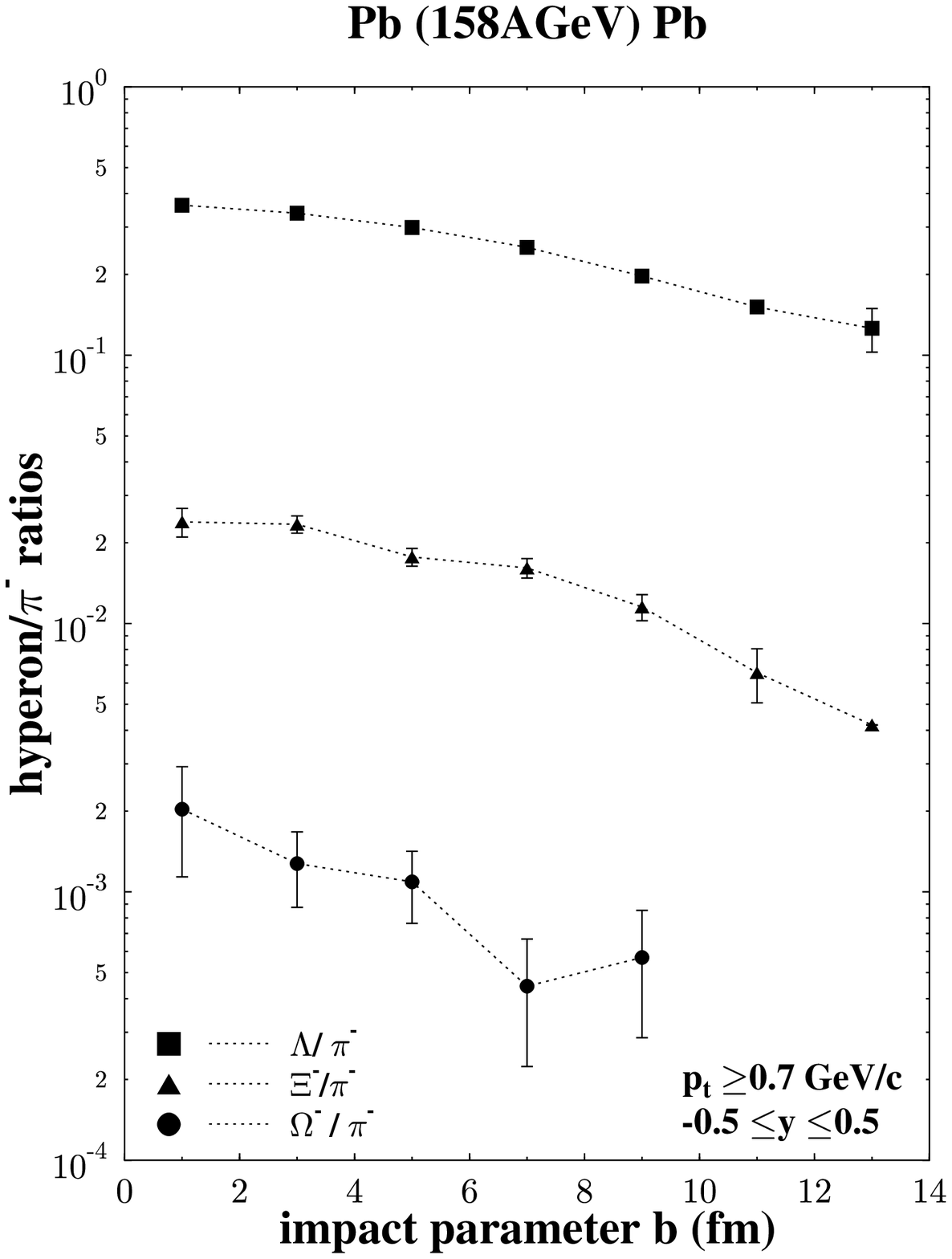,width=8cm}} 
\vspace{10pt}
\caption{ 
Hyperon to $\pi^-$ ratio as a function of impact parameter $b$, 
as obtained from the three-fluid hydrodynamical model (Left) 
and the UrQMD model (Right). In the UrQMD model, 
the observed strangeness enhancement is already a natural consequence 
of ordinary hadronic rescattering.} 
\label{urqmd_ratios}
\end{figure}
\begin{figure}[thb]
\centerline{\hbox{\epsfig{file=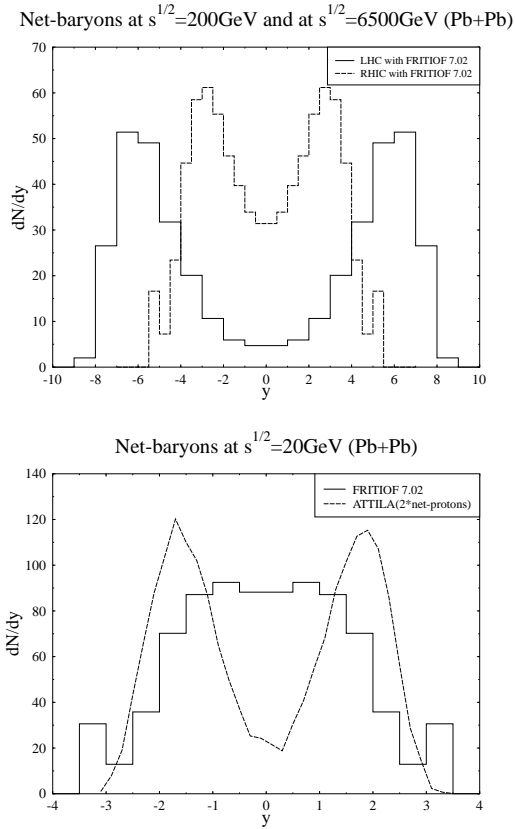,height=4.8in}}}
\vspace{10pt}
\caption{
Net-baryon rapidity distribution of very central $Pb+Pb$ collisions 
at SPS, RHIC, LHC calculated with FRITIOF 7.02. 
The midrapidity region is even at LHC not net-baryon-free. 
For comparison the net-protons at SPS calculated with ATTILA are also shown.}
\label{ger1}
\end{figure}  

Let us now turn to multi-strange signals. In nucleon nucleon collisions, 
the production of particles containing strange quarks is strongly suppressed 
as compared to the production of particles with $u$ and $d$ quarks 
due to the higher mass of the $s \bar{s}$ quark pair.

It has been speculated that the yield of strange and multi-strange mesons,
(anti-) baryons and anti-hyperons 
($\bar{\Lambda}, \bar{\Sigma}, \bar{\Xi}$ and $\bar{\Omega}$) should be 
enhanced in the presence of a QGP.

The study of (multi)strange hyperons by the WA97 \cite{lietava} and 
the NA49 collaborations show an enhancement 
of strangeness production for central collisions when studying 
the centrality dependence of various strange particle yields 
($\Lambda, \Xi, \Omega$) in $Pb+Pb$ collisions at 158 AGeV as compared
to $p+Pb$ collisions at 158 AGeV. The centrality is given as the 
extrapolated number of participant nucleons $N_{\rm part}$. 
We propose as centrality variable the number of produced pions $N_{\pi^-}$.
$N_{\rm part}$ shows a nonlinear behavior with the volume of the 
participant zone, while $N_{\pi^-}$ shows perfect participant scaling. 
Scaling has been observed for central collisions 
($N_{\rm part} \geq 100$).

The UrQMD calculations (Fig.~\ref{urqmd_ratios}, right) show scaling. 
The hyperon to $\pi^-$ ratio is depicted in Fig.~\ref{urqmd_ratios} (Right)
as a function of impact parameter $b$. For central collisions, 
all ratios change only moderately, thus an approximate 
linear scaling of the hyperon yield with pion number $N_{\pi^-}$ is observed. 
For peripheral collisions, the ratios decrease. 
The ratios vary with a factor of 2 to 5 for different impact parameters 
depending on the hyperon and its strangeness content.
The three-fluid hydrodynamical model with an EoS with
a first order phase transition to a QGP yields constant ratios 
(Fig.~\ref{urqmd_ratios} left). Note the substantial differences in
the $\Xi/\pi$-ratios between the two predictions.
A comparison to upcoming data by the NA49 and CERES may provide an
estimate of the degree of local chemical equilibration. 

\begin{figure}[t!]
\centerline{\epsfig{file=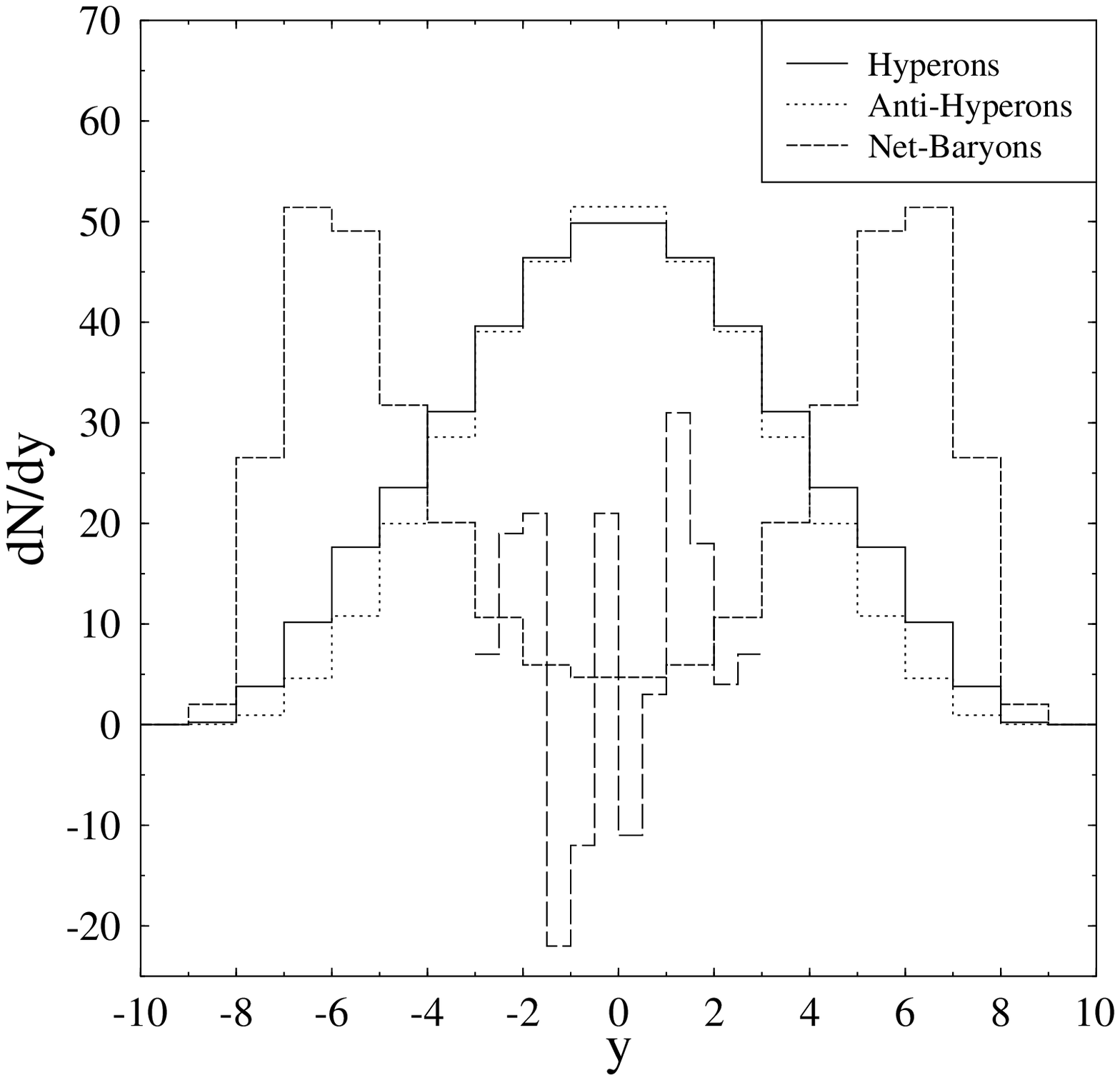,width=8cm}
\hfill
\epsfig{file=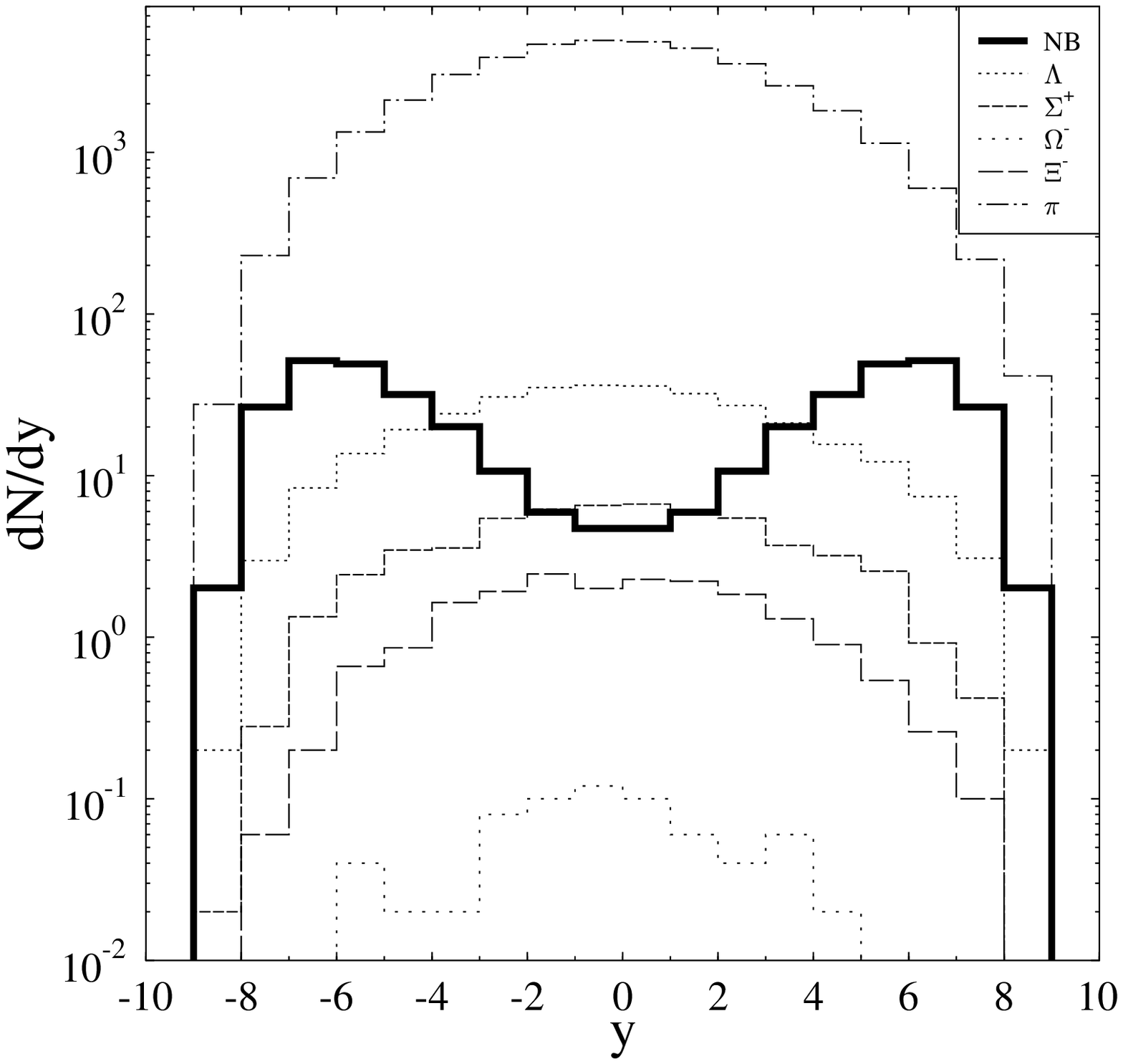,width=8cm}}
\vspace{10pt}
\caption{
(Left) The (anti-)hyperon rapidity distribution and mean net-baryon 
distribution at midrapidity compared with the distribution of a single event,
and 
(Right) Rapidity distributions of different particles 
(NB denotes net baryons), both in very central $Pb+Pb$ collisions at 
$\protect\sqrt{s}_{NN}=6.5$ TeV,  
calculated with FRITIOF~7.02.}
\label{ger2}
\end{figure}
\begin{figure}[t!]
\centerline{\hbox{\epsfig{file=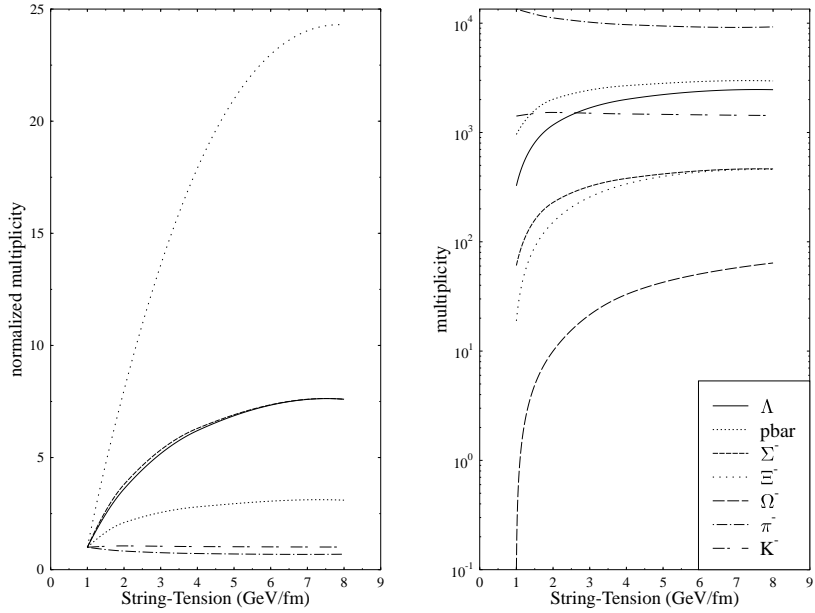,width=11cm}}}
\vspace{10pt}
\caption{ 
The multiplicities of different particles in very
central $Pb+Pb$ collisions at LHC calculated with FRITIOF~7.02 as
function of the string-tension.}
\label{ger4}
\end{figure}

Fig.~\ref{ger1} exhibits the baryon rapidity distribution as predicted by 
various models for heavy ion collisions. 
ATTILA~\cite{attila} and FRITIOF~1.7~\cite{frit} (not in the picture) 
show nearly a baryon-free midrapidity region already at SPS(CERN). 
These models are therefore ruled out by the new CERN data,
which rather support predictions based on the RQMD model~\cite{sorge}.
Also the new Lund model release FRITIOF~7.02 yields stopping at SPS! 
At RHIC, FRITIOF~7.02 and RQMD~\cite{tomi} predict that the net baryon number
$A \gg 0$ at $y_{\rm cm}$.
Furthermore, even in very central collisions of lead on lead at
$\sqrt{s}_{NN}=6.5$~TeV, there might be some net-baryon density 
at midrapidity. This is shown in Fig.~\ref{ger2} (Left), where the event-averaged 
rapidity densities of net-baryons, hyperons and anti-hyperons are 
depicted for LHC, using FRITIOF~7.02.
If this non-perfect transparency turns out to be true, the finite 
baryo-chemical potential at midrapidity may have strong impact 
on the further evolution of the system. Expected yields of strangelets will 
be extremely sensitive to the initial baryon-number of a 
Quark-Gluon-Plasma-phase.

Fig.~\ref{ger2} (Right) shows the event-averaged rapidity densities of net-baryons, 
hyperons and pions calculated with FRITIOF~7.02. Note that the strange 
to non-strange hadron ratios predicted by this model are the 
same for $pp$ and $AA$ collisions at 200 AGeV/c (CERN-SPS) and that the 
strange particle numbers for $AA$ underpredict the data~\cite{antai}.
This deficient treatment of the collective effects in the model leads us
to take the numbers only as lower limits of the true strange particle   
yields at collider energies.

Keep in mind that the microscopic models used here ignore possible
effects that could change significantly the number of produced 
strange particles in heavy ion collisions, 
e.g. string-string-interactions.
An enhanced string tension may effectively simulate
string-string interaction, as shown in Fig.~\ref{ger4}.
Here the multiplicities of different produced particles at LHC as 
function of the string-tension $\kappa$ are depicted.
A higher string-tension, e.g. 2 GeV/fm yields
the suppression factors \quad u : d : s : qq =  1 : 1 : 0.55 : 0.32.

%
%

\section*{Dilepton production}

\begin{figure}[!t] 
\centerline{\epsfig{file=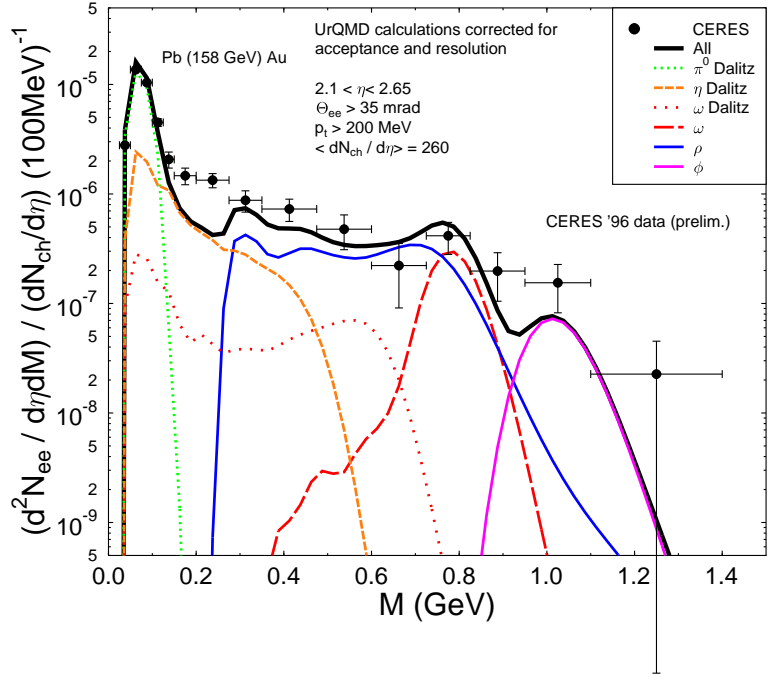,width=11cm}} 
\vspace{10pt}
\caption{
Microscopic calculation of the dilepton production 
in the kinematic acceptance region of the CERES detector 
for $Pb+Au$ collisions at 158~GeV. No in-medium effects are taken 
into account. Plotted data points are taken at CERES in '96.} 
\label{dilepton_pbau}
\end{figure}
\begin{figure}[!t] 
\centerline{\epsfig{file=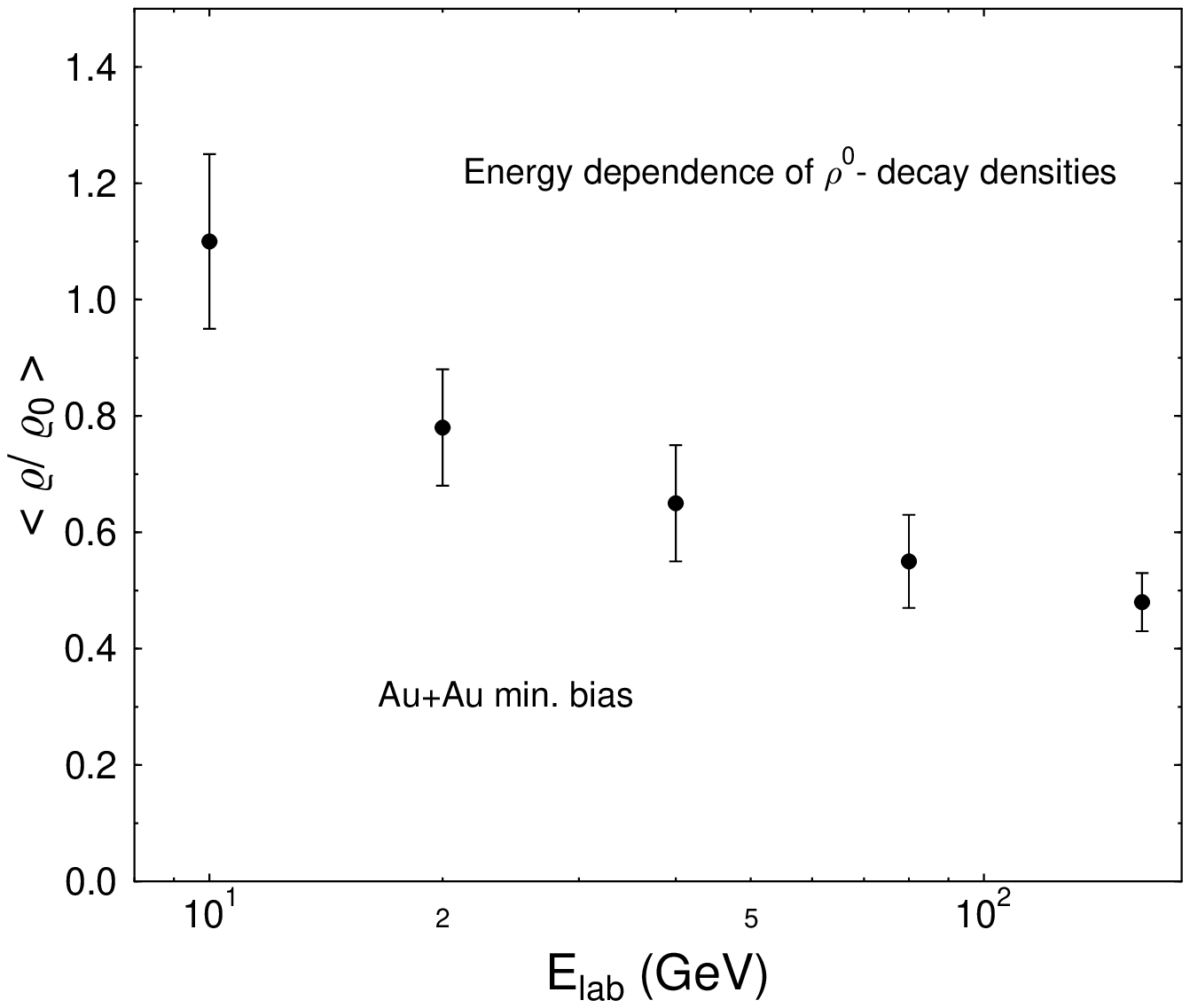,width=8cm}
\hfill 
\epsfig{file=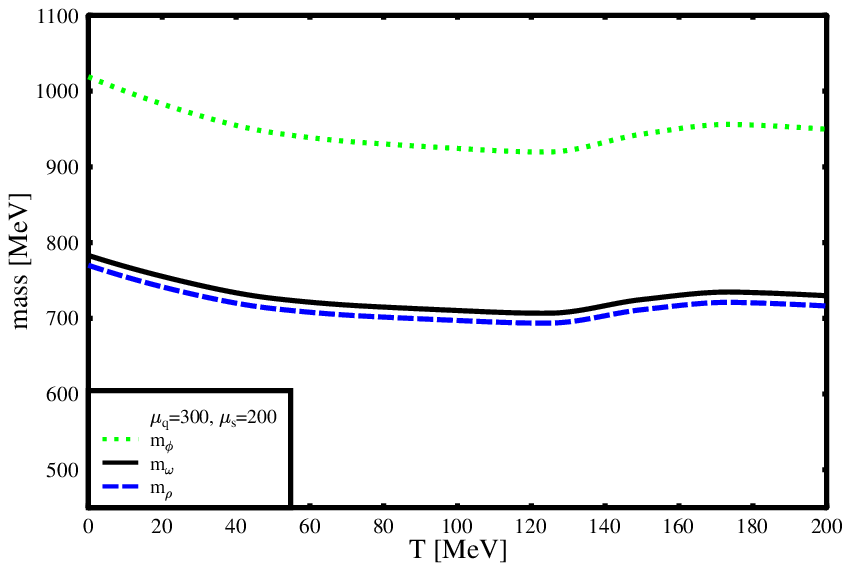,width=8cm}} 
\vspace{10pt}
\caption{
(Left) The mean ``freeze-out'' density 
at the location of $\rho$ meson decays in $Au+Au$ collisions. 
(Right) The mass of the $\rho$-meson as 
obtained from the chiral model at different temperatures and finite density.} 
\label{dilepton_energy} 
\end{figure}

Besides results from hadronic probes, electromagnetic radiation 
-- and in particular dileptons -- 
offer an unique probe from the hot and dense reaction zone: 
here, hadronic matter is almost transparent. 
The observed enhancement of the dilepton yield 
at intermediate invariant masses ($M_{e^+e^-} > 0.3$~GeV) 
received great interest: it was prematurely thought that
the lowering of vector meson masses is required by chiral symmetry restoration 
(see e.g. \cite{koch97} for a review). 
However, there seems no theoretical support for this speculation.
Calculations within a chiral $SU(3)$ mean-field approach \cite{papa98} 
show only a modest dependence of temperature of the mass of the $\rho$~meson
(Fig.~\ref{dilepton_energy}, right). AA-data are compatible 
with broadening spectral functions found in pure hadronic calculations 
of the scattering on the constituents of the excited matter 
(see e.g.~\cite{cassing97a}). 
The present data do not allow to draw definite conclusions. 

Fig.~\ref{dilepton_pbau} shows a microscopic UrQMD
calculation of the dilepton production in the kinematic 
acceptance region of the CERES detector for $Pb+Au$ collisions at 158~GeV. 
This is compared with the '96 CERES data \cite{agakishiev97b}. 
Aside from the difference at $M\approx 0.4$~GeV there is a strong 
enhancement at higher invariant masses. 
It is expected that this discrepancy at $m>1$~GeV could be filled up 
by direct dilepton production in meson-meson collisions \cite{liG98a} 
as well as by the mechanism of secondary Drell-Yan 
pair production proposed in \cite{spieles97a}.

The mean ``freeze-out'' density at the location of $\rho$ meson decays 
in $Au+Au$ collisions is shown in Fig.~\ref{dilepton_energy} (Left)
for different incident energies \cite{winckelmann96b}. 
From AGS to CERN energies, there is a decrease of the baryonic density, 
indicating that baryonic modifications to the $\rho$ meson 
are better studied at energies of $20-40$~AGeV. 
The low baryon densities at high energies will make it hard 
to explain the CERES data by $\rho$ meson modifications 
of nucleonic origin alone.

%
%
%

\section*{Collective Flow and the softening of the EoS}

The excitation function of collective transverse flow 
is the earliest predicted signature 
for probing compressed nuclear matter. 
Transverse collective flow depends directly on the pressure $p(\rho,S)$,
i.e.\ the EoS. The flow excitation function is sensitive to 
phase transitions \cite{stoecker86a} by a collapse of the directed 
transverse flow \cite{scheid76,ha90}. This is commonly referred to as 
{\em softening} of the EoS \cite{rischke95b}. 

\begin{figure}[b!] 
\centerline{\epsfig{file=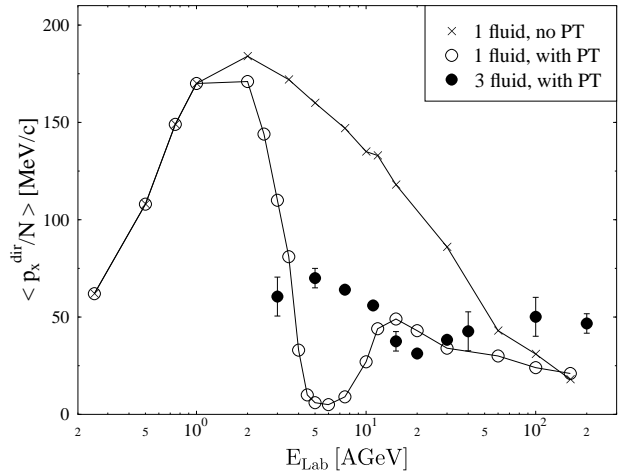,width=10cm}} 
\vspace{10pt}
\caption{
The excitation function of transverse flow as 
obtained from one fluid hydrodynamics with (open circles) 
and without (crosses) a first order phase transition 
\protect\cite{rischke95b}, and the results of the three-fluid hydrodynamical 
model (filled circles). 
The drop of $p_x$ due to the softening of the EoS is shifted to 
$E_{\rm lab}\approx20$~AGeV.} 
\label{hydro_flow} 
\end{figure}

An observation of a local minimum in the excitation function 
of the transverse directed flow would thus be an unambiguous signal 
for a first order phase transition in dense matter. 
It's experimental measurement would serve as strong evidence for a QGP, 
if that phase transition is of first order.

\begin{figure}[tbh] 
\centerline{\epsfig{file=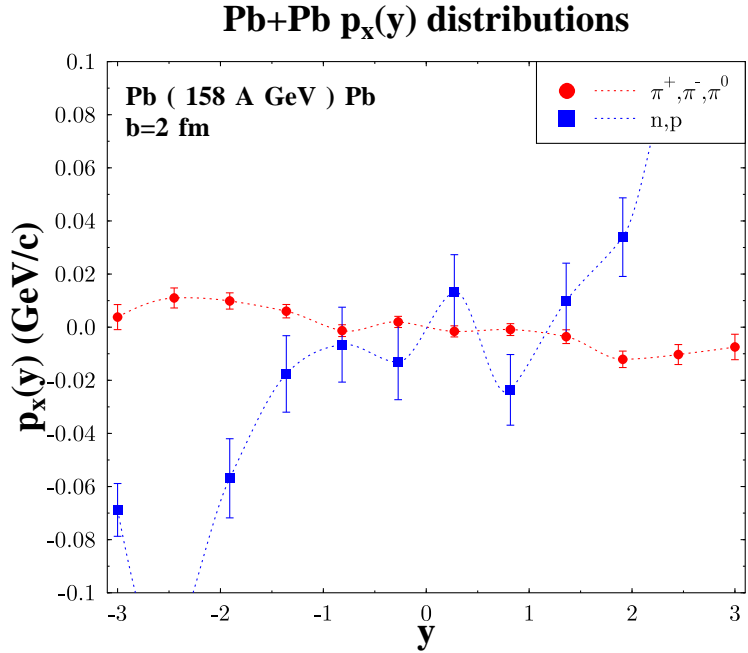,width=10cm}} 
\vspace{10pt}
\caption{
Theoretical predictions for transverse collective
flow in $Pb+Pb$ collisions as obtained with the microscopic UrQMD transport
model \protect\cite{bleicher96}.} 
\label{flow_soff} 
\end{figure}

Recent calculations within three-fluid hydrodynamics \cite{brachman98} 
show a shift in the drop of transverse flow to higher energies, 
$E_{\rm lab}\approx 20$~AGeV, see Fig.~\ref{hydro_flow}.
Experimentally, the recent discovery of proton flow and pion antiflow 
at the SPS is in line with UrQMD and RQMD predictions (Fig.~\ref{flow_soff}, see
\cite{keitz} and \cite{bleicher96}).

%
%

\section*{Outlook}

At the CERN/SPS new data on flow, electro-magnetic probes, 
strange particle yields (most importantly multistrange (anti-)hyperons) 
and heavy quarkonia will be interesting to follow closely. 
Simple energy densities estimated from rapidity distributions 
and temperatures extracted from particle spectra indicate 
that initial conditions could be near or just 
above the domain of deconfinement and chiral symmetry restoration.
Still the quest for an {\em unambiguous} signature remains open.

Directed flow has been discovered 
-- now a flow excitation function, filling the gap between 10 AGeV (AGS) 
and 160 AGeV (SPS), would be extremely interesting to look for the 
softening of the QCD equation of state in the coexistence region. 
The investigation of the physics of high baryon density 
(e.g. partial restoration of chiral symmetry via properties 
of vector mesons) is presently not accessible due to the 
lack of dedicated accelerators in the $10-200$ AGeV regime. 

However, dedicated accelerators would be mandatory to explore these 
intriguing effects in the excitation function. 
It is questionable whether this key program will actually get 
support at CERN. Also the excitation function of particle yield 
ratios ($\pi/{\rm p}, {\rm d}/{\rm p}, {\rm K}/\pi ...$) and, in particular, 
multistrange (anti-)hyperon yields, can be a sensitive probe of 
physics changes in the EoS.
The search for novel, unexpected forms of $SU(3)$ matter, e.g. {\em hypermatter}, 
{\em strangelets} or even {\em charmlets} is intriguing. Such exotic QCD 
multi-meson and multi-baryon configurations would extend 
the present periodic table of elements into hitherto 
unexplored dimensions. A strong
experimental effort should continue in that direction.

For applications to nuclear collision observables,
an extension of the QGP concept to non-equilibrium conditions
is required. The popular use of simple fireball
models may provide convenient parameterizations
of large bodies of data, but they will never provide
a convincing proof of new physics. 
Microscopic transport models are required that can address
simultaneously all the observables and account for experimental 
acceptance and trigger configurations. 

Present work in parton cascade dynamics is based largely on analogy 
to transport phenomena in known abelian QED plasmas. 
A significant new feature of QCD plasmas
is its ultrarelativistic nature and the dominance of (gluon) radiative  
transport. These greatly complicate the equations.
The role of quantum coherence phenomena beyond classical
transport theories has only recently been established
within idealized models. Much further work will be required
in this connection. The outstanding theoretical task
will be the development of practical (vs. formal)
tools to compute quantum non-equilibrium
multiple collision dynamics in QCD. 

Experiments and data on ultra-relativistic collisions 
are essential in order to motivate, guide, and constrain 
theoretical developments. They provide the only terrestrial 
probes of non-perturbative aspects of QCD and its dynamical vacuum. 
The understanding of confinement and chiral symmetry remains one of 
the key questions at the beginning of the next millennium.

\section*{Acknowledgments}

This work was supported by DFG, GSI, BMBF, DAAD, Graduiertenkolleg Theoretische 
und Experimentelle Schwerionenphysik, the A.~v.~Humboldt Foundation, 
and the J.~Buchmann Foundation.

This paper is dedicated to our colleague and friend, Klaus Kinder-Geiger. 
Over the last decade, we have had the great pleasure of sharing frequent,
sometimes heated, scientific discussions about the physics and the future 
of our field with Klaus.

Klaus, Du fehlst uns.

\end{document}